\documentclass[a4paper]{spie}  

\usepackage[]{graphicx}
\def\aaomega{AAOmega}
\def\etal{\emph{et\,al.}}

\title{Performance of AAOmega: the AAT multi-purpose fibre-fed
spectrograph.}

\author{
Robert Sharp$^{\rm{1}}$,
Will Saunders$^{\rm{2}}$,
Greg Smith,
Vladimir Churilov,
David Correll,
John Dawson,
Tony Farrel,
Gabriella Frost,
Roger Haynes,
Ron Heald,
Allan Lankshear,
Don Mayfield,
Lew Waller,
Dennis Whittard
\skiplinehalf
Anglo-Australian Observatory, P.O. Box 296, Epping NSW 1710, Australia
}

\authorinfo{Send correspondence to $^{\rm{1}}$Rob Sharp:
rgs@aao.gov.au, $^{\rm{2}}$Will Saunders: will@aao.gov.au}

\begin{document} 
\maketitle


\begin{abstract}
AAOmega is the new spectrograph for the 2dF fibre-positioning system
on the Anglo-Australian Telescope. It is a bench-mounted,
double-beamed design, using volume phase holographic (VPH) gratings
and articulating cameras. It is fed by 392 fibres from either of the
two 2dF field plates, or by the 512 fibre SPIRAL integral field unit
(IFU) at Cassegrain focus. Wavelength coverage is 370 to 950nm and
spectral resolution 1,000-8,000 in multi-Object mode, or 1,500-10,000
in IFU mode. Multi-object mode was commissioned in January 2006 and
the IFU system will be commissioned in June 2006.

The spectrograph is located off the telescope in a thermally isolated
room and the 2dF fibres have been replaced by new 38m broadband
fibres. Despite the increased fibre length, we have achieved a large
increase in throughput by use of VPH gratings, more efficient coatings
and new detectors - amounting to a factor of at least 2 in the
red. The number of spectral resolution elements and the maximum
resolution are both more than doubled, and the stability is an order
of magnitude better.

The spectrograph comprises: an f/3.15 Schmidt collimator,
incorporating a dichroic beam-splitter; interchangeable VPH gratings;
and articulating red and blue f/1.3 Schmidt cameras. Pupil size is
190mm, determined by the competing demands of cost, obstruction
losses, and maximum resolution. A full suite of VPH gratings has been
provided to cover resolutions 1,000 to 7,500, and up to 10,000 at
particular wavelengths.
\end{abstract}



\section{Introduction}
\label{sect:intro}  
The two degree field facility (2dF$^{\rm{[1,2]}}$) of the
Anglo-Australian Telescope (AAT), had its original spectrographs
mounted on the telescope top end ring. This was mandated by the
capabilities of fibres at that time, and the spectrographs were
designed for the primary science goal - a large redshift survey of
moderate brightness galaxies. However, they were of limited utility in
obtaining spectra of fainter objects, or at higher signal-to-noise or
resolution. The reasons for this were:
\begin{itemize}
\item Restricted size and weight, with implications for instrument
stability and format
\item Flexure and temperature variations, compromising both spectral
and spatial stability of the system
\item Imperfect optics giving variable point spread function (PSF),
leading to systematic error in skyline subtraction that could not be
reduced by increased integration.
\item Significant scattered light and halation problems, limiting the
dynamic range within and between spectra.
\end{itemize}

Despite these limitations, both spectrographs performed admirably
resulting in a multitude of high quality scientific results over their
10 year careers.  In August 2005 spectrographs 1 and 2 were retired
from service to make way for installation of the next generation AAT
multi-purpose optical spectrograph, \aaomega.

The mechanical$^{\rm{[3]}}$ and optical$^{\rm{[4]}}$ design of
\aaomega\ has been presented previously elsewhere.  In what follows we
briefly describe the \emph{as commissioned} capabilities of \aaomega.
The commissioning program is outlined in table \ref{tab:time line}.

\begin{table}[h]
\caption{\aaomega\ was commissioned November 2005 - January 2006
following the project time line outlined below.}
\label{tab:time line}
\begin{center}       
\begin{tabular}{|l|l|}
\hline
\textbf{Date} & \textbf{Status}\\
\hline
\hline
August 2005 & Decommission 2dF spectrographs\\
            & Remove 2dF fibre retractors to install \aaomega\ fibres\\
\hline
November 2005 & First on sky commissioning (6 nights)\\
              & 2dF positioner upgrade and new field plates\\
              & Instrument and telescope control software integration\\
              & AAOmega guide fibres\\
              & Limited AAOmega slit units\\
              & (2 Plates $\times$ 3  Bundles $\times$ 10 fibres)\\
              & Blue camera only\\
\hline
December 2005 & Second commissioning run (7 nights)\\
              & Red and Blue cameras\\
              & Fully populated slit units\\
\hline
January 2006  & Final commissioning run (5 nights)\\
              & Full system integration\\
\hline
\hline
January 2006  & Science verification (8 nights)\\
              & 9 programs substantially completed\\
              & Testing the full range of \aaomega\ capabilities.\\
\hline
February 2006  & Commence routine science operations\\
 - May 2006   & 10 scheduled programs, 41 nights\\
              & Active service program\\
\hline
\hline
June 2006     & SPIRAL IFU Scheduled for commissioning\\
\hline
\end{tabular}
\end{center}
\end{table}

\subsection{Improvements over the original 2dF system}
While maintaining the high multiplex (392 fibres) and low
observational overhead (via configuration of one field plate during
observation of the other) of 2dF, the key design goals for \aaomega\
have been to overcome many of the problems experienced with the
original 2dF system:

\begin{itemize}
\item Improved sensitivity at all wavelengths, even with the increased
fibre run length.
\item Spectroscopic stability - removal of thermal and flexure effects
\item PSF uniformity - dramatically improved sky subtraction and
higher velocity accuracy
\item Higher maximum resolution
\item Greater wavelength coverage at a given resolution
\item Minimisation of systematic effects, such as scattered light and
light leaks
\item Improved support for \emph{Nod-and-Shuffle} observations$^{\rm{[5]}}$.
\end{itemize}

As an alternate to the 2dF MOS fibre feed, the SPIRAL Integral Field
Unit (IFU) mounted at the Cassegrain focus, will also be available to
\aaomega.  SPIRAL will be commissioned in June 2006 and hence we defer
discussion of the IFU feed until a later time.

\section{Details of the final implementation of AAOmega systems.}

\subsection{Fibre positioner and fibre run}
\aaomega\ retains much of the 2dF fibre positioner system at the prime
focus of the AAT and hence we refer the reader to the previous
review$^{\rm{[2]}}$ for a detailed review of 2dF. The major departures
from the original system are the replacement of the 8m fibre run with
the 38m run required to deliver the fibre slit units from prime focus
to the Coud\'e west laboratory to feed the bench mounted dual beam
\aaomega\ spectrograph.  A continuous fibre run design was considered
preferable to a system with a fibre connector, to eliminate light
losses and focal ratio degradation at a connector.  The fibre run
remains permanently attached to 2dF and is routed to the spectrograph
via the Serrurier truss, the Coud\'e tunnel and the horseshoe bearing
before arriving at Coud\'e west. The Polymicro Broadband
fibres\footnote{http://www.polymicro.com/products/opticalfibers/products\_opticalfibers\_fbp.htm}
have a nominal core diameter of 140um, cladding diameter 168um and
polyimide buffer diameter of 196um. Measured throughput is 72\% at
400nm and 96\% at 800nm.

\subsection{Guide bundles}
Each 2dF field plate was originally equipped with four \emph{coherent}
bundles, constructed from a close packed arrangement of seven fibres.
For \aaomega\ an additional four guide bundles per field plate were
added, greatly increasing acquisition accuracy in normal use. It also
allows four guide stars to be acquired in each telescope position of a
Nod+Shuffle cross-beam-switching pair (see section \ref{sec:sky}).

\subsection{The spectrograph}
The spectrograph design has previously been described$^{\rm{[3,4]}}$;
we here focus on the comparison between design and in-use performance.

The design has a 145mm fibre slit, with a field lens to make it
telecentric (or more correctly, pupil-centric); an f/3.15 Schmidt
collimator incorporating a dichroic beam-splitter upstream of the
corrector (this necessitates two correctors but minimises pupil
relief); in each arm there is then a collimator Schmidt corrector; an
articulating VPH grating; then an f/1.3 Schmidt camera with a doublet
corrector, spherical mirror, a field-flattening lens of S-YGH51 glass,
and a 2K x 4K E2V CCD with the long axis in the spatial
direction. Optics, coatings and detectors were all specified
separately for each arm.

The design emphasised excellent optics and image quality, to give a
uniform PSF and hence good skyline subtraction. The most difficult
components were the 230mm diameter aspheric doublet corrector plates
for the wide-field (8$^{\circ}$ off-axis angles), fast (f/1.3) Schmidt
cameras. These were made by SE Laser Physics St Petersburg. The
overall surface quality was close to the half-wave specification, but
small scale radial gradient errors were a concern.

\begin{figure}
  \begin{center}
    \begin{tabular}{c}
      \includegraphics[height=9cm]{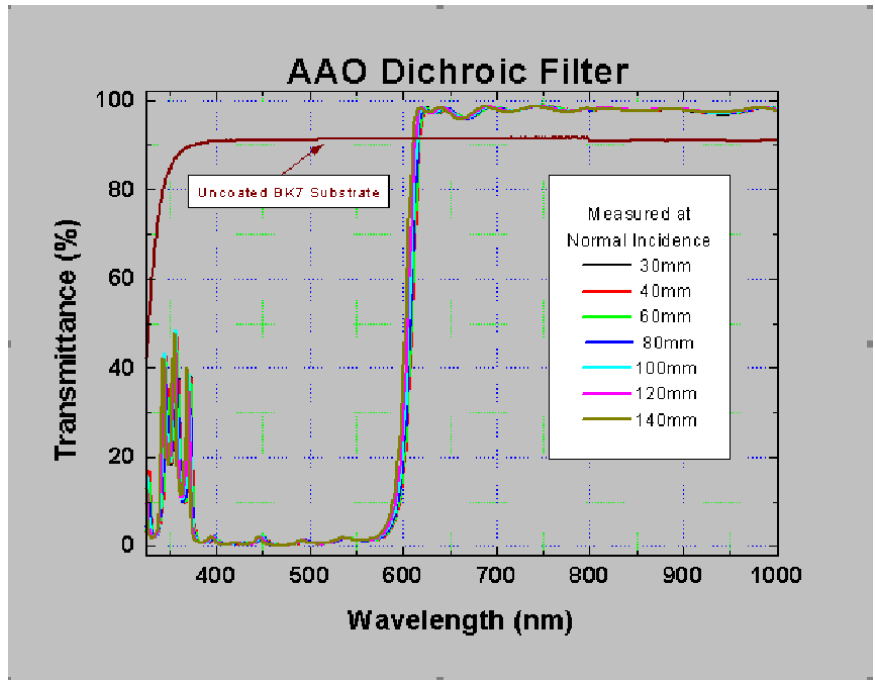}
    \end{tabular}
  \end{center}
  \caption{\label{fig:dichroic} Response of the 570nm dichroic
  currently available for use with AAOmega.  Note the abrupt
  transition region and generally smooth profile.}
\end{figure}

The dichroic beam-splitter was manufactured by Precision Optics at
CSIRO Lindfield, and meets the very demanding specification (figure
\ref{fig:dichroic}). Only a single beam-splitter, with crossover at
570nm, is currently in use, but others are envisaged as needed for
particular science programs.  The present cross over wavelength is
chosen to give continuous wavelength coverage, at low resolution, from
$\sim$370nm to $\sim$880nm, while still placing the strong blue sky
line at 5577\AA\ within the coverage of the blue spectrum.  The line
is used to allow relative fibre throughput calibration, a process
required for sky subtraction using dedicated sky fibres (see section
\ref{sec:sky}), avoiding the need for costly offset sky calibrations.

The VPH gratings were made by Ralcon Development Lab. The \emph{blue}
($<$570nm) gratings used commercial Starphire glass, post-polished to
give half-wave image quality; the \emph{red} gratings used
pre-polished BS7 substrates. All gratings are 20mm thick. At the time
of procurement we were not able to locate good quality affordable
BK7/UBK7 substrates; this situation has now changed
(e.g. www.infiniteoptics.com).

\begin{figure}
  \begin{center}
    \begin{tabular}{c}
      \includegraphics[height=6cm]{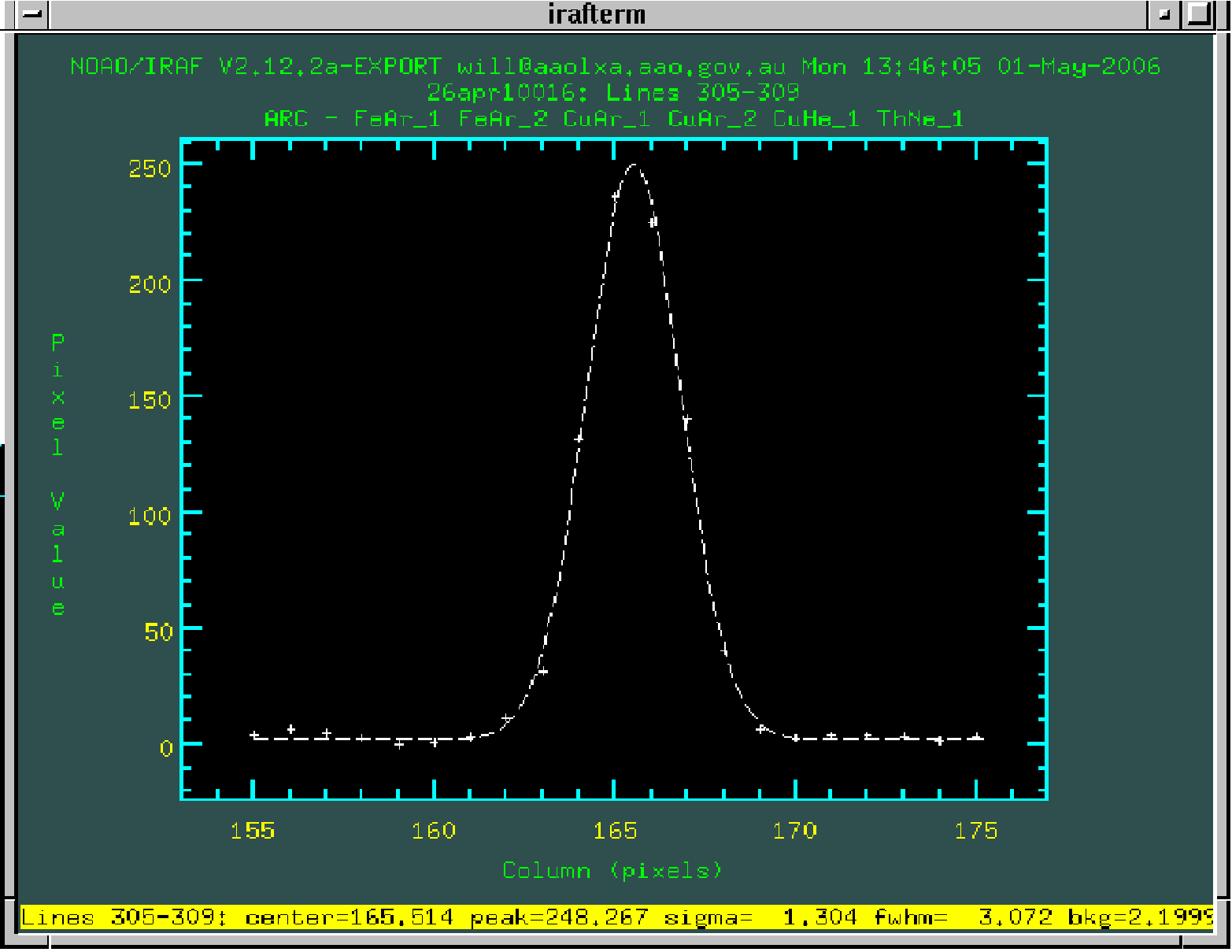}
      \includegraphics[height=6cm]{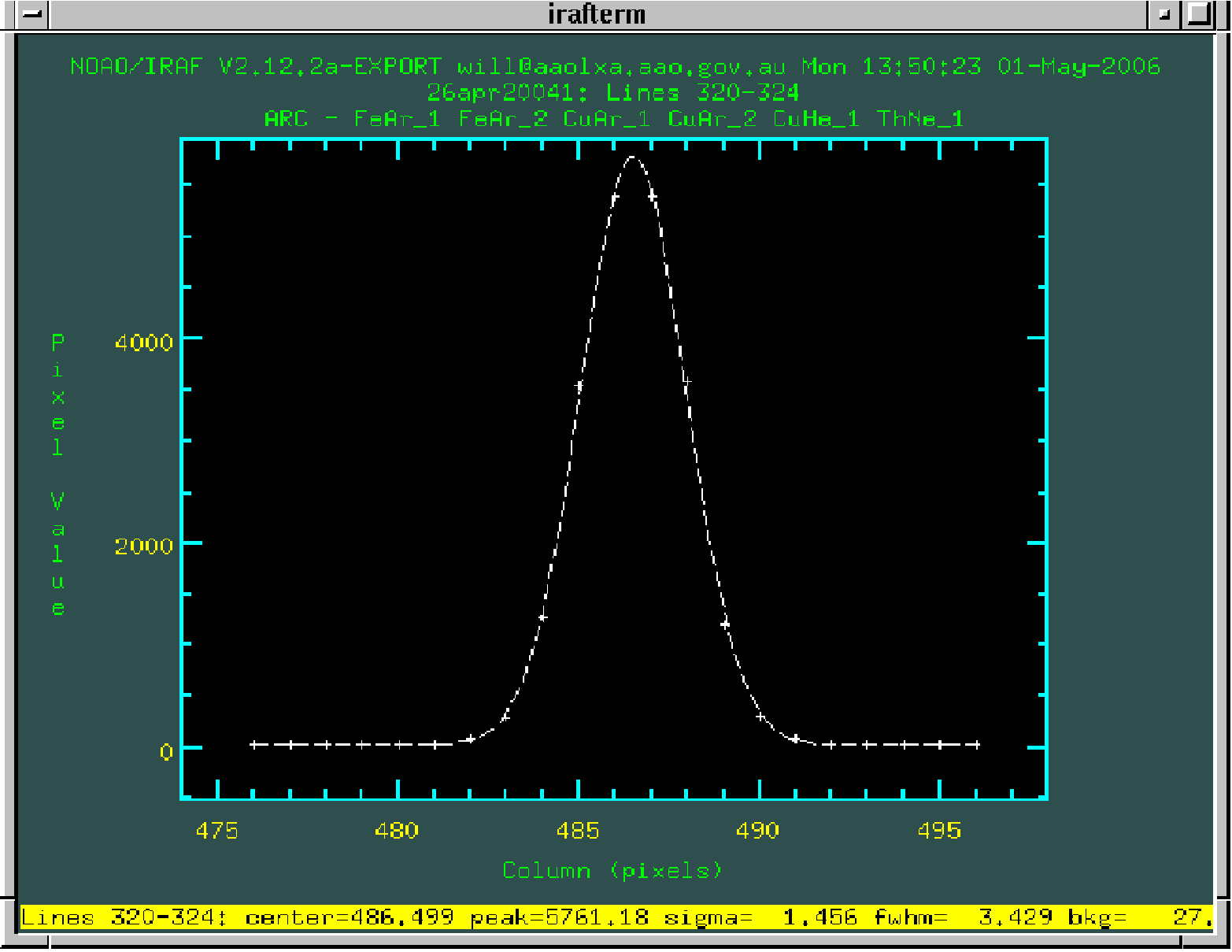}
    \end{tabular}
  \end{center}
  \caption{\label{fig:PSF} Examples of the blue (left) and red (right)
  fibre PSF show the system to be satisfactory, and highly Gaussian.
  The poorer red performance is believed to have its origins in charge
  diffusion in the deep-depletion CCD.}
\end{figure}

The optical design gave a theoretical PSF which was closely Gaussian,
with a full width at half maximum (FWHM) of 3.2 pixels, and a
variation of less than a few percent over the detector. The actual,
in-use PSF is indeed almost perfectly Gaussian (figure
\ref{fig:PSF}). The blue camera has a FWHM and uniformity as good as
specified, with actual FWHM in the range 3.1-3.3 pixels. The red
camera is less satisfactory, giving 3.3-3.6 pixels, but extremely
uniform for any particular wavelength.

The poorer red focus was surprising in light of the generally easier
optics for the red camera.  The strongly suspected explanation is the
combination of fast optics with the deep-depletion (DD) CCD. Our
results suggest an extra contribution from charge diffusion of order 1
pixel (15$\mu$m) for the red camera. The DD silicon thickness is
40$\mu$m, vs 16$\mu$m for the blue arm CCD, and the typical
penetration of the photons will be proportionately larger. This leads
to increased charge diffusion even for on-axis incoming light, but the
effect is compounded for a fast beam such as ours. In practice, the
poorer than expected red camera focus leads to no problems except a
10\% lower resolution than anticipated.

Scattered light performance has been excellent, with 5\% (red, average
at low dispersion) and 23\% (blue) of the overall light incident on
the detectors not being extractable into spectra.  This results from
the relatively small number of optical surfaces in the spectrograph
(13) and the excellent performance of the gratings themselves.  The
higher scattering in the blue camera is due to residual ``moisture''
contamination from initial assembly and is improving with time (see
section \ref{sec:frost}).

\subsection{The cameras}
The overall camera design remains the same as in previous discussions
of AAOmega$^{\rm{[3]}}$.  During routine operations vacuum hold times
have proven longer than intervals between scheduled interventions in
their operation as part of the commissioning process ($\sim$3\,weeks).
Achieving the detector operating temperature of 160K in the Epping
laboratory prior to installation at the AAT, proved difficult, a
potential problem for the blue camera due to the intrinsic dark
current.  Further investigation showed this to be due to the radiative
thermal load on the detector, through the camera optics, in the warm
clean room.  There has been little problem achieving the required
set-point once installed at the AAT.

\section{AAOmega in use}

\subsection{Acquisition and guidance}

The guide fibres are filtered at $\lambda$5000\AA\ and imaged with a
Watec CCD camera giving improved intensity control and sensitivity
over the older 2dF Quantex video system.  The preferred guide star
magnitude range is currently \emph{V}=14-14.5\,mag(Vega) (although for
all practical purposes \emph{B}, \emph{V} or \emph{R} band limits are
interchangeable here) with 15\,mag representing the lower practical
limit.  During bright time observations, 12\,mag stars are more
practical.  The Watec camera has a limited dynamic range, and hence
the spread in guide star magnitude during any individual observation
should be constrained to be $\Delta$mag$<$0.5 for the best results.
An auto-guider system provides translational motion on the 1-5\,sec
scale, and also estimates any field plate rotation corrections. The
latter is required both to correct for small amounts of
declination-dependent flexure in the 2dF system, and also to
(partially) correct for changes in differential atmospheric refraction
as the field tracks across the sky.

\subsection{Detector modes}
AAOmega is a general purpose instrument providing many modes of use.
Windowing and spectral binning are supported. Each detector can be
read out through one or both amplifiers at a variety of read
speeds. We have achieved read noises of 3.8e$^{-}$ and 4.5e$^{-}$ at
\textsc{normal} (125s full frame single amplifier readout time) and
\textsc{fast} (74s) read speeds.

\subsection{Astronomical use of VPH gratings}
VPH gratings$^{\rm{[6]}}$ offer great advantages for astronomy over
traditional reflection gratings. They are cheaper, more efficient, can
be designed to any desired specification, can be made in large
formats, have less scattered light, offer much better pupil relief,
and there is now a three-way choice of manufacturer. The drawbacks are
that the response is more strongly peaked, and that they do not work
well at very low resolutions. In several respects, they differ greatly
in use from reflection gratings:

\begin{itemize}
\item The wavelength coverage for a particular grating angle is
determined by the camera angle, which must hence be adjustable
\item The peak efficiency wavelength can be altered by adjusting the
grating angle; this has secondary effects on the wavelength coverage
and resolution
\item In any given setup, the resolution is more strongly wavelength
dependent than for reflection gratings
\end{itemize}

Since there is an extra degree of freedom over reflection gratings,
great care is needed in setting up the observations in the most
efficient way. To help with this, a grating
calculator\footnote{http://www.aao.gov.au/cgi-bin/aaomega\_calc.cgi}
has been provided. All transmission gratings used in Littrow mode
suffer from a 0$^{\rm{th}}$ order ghost image of the slit, caused by
dispersed light reflected off the CCD, recollimated by the camera
optics, diffracted in 1$^{\rm{th}}$ order mode back into an
undispersed beam, and reimaged onto the camera. The ghost typically
amounts to a few percent in intensity compared with the actual
spectra. This ghost can be moved off the detector by tilting the
grating, and hence breaking the Littrow symmetry, and most of the
AAOmega gratings have their internal refractive index fringes slanted
(like a Venetian blind) to allow this symmetry breaking without
damaging the efficiency characteristics.

\subsection{\label{sec:sky}Sky subraction modes}
The single greatest limitation for fibre-fed spectroscopic data is
accurate sky subtraction. For precision stellar work, accurate
subtraction of the solar spectrum is essential; for faint object red
spectroscopy, a forest of OH telluric skylines dominates the sky
spectrum and must be subtracted precisely. AAOmega offers a variety of
methods for sky subraction, offering a trade off between this
systematic noise vs statistical noise and number of targets that can
be observed.

The most common mode is sky subtraction with \emph{dedicated sky
fibres}. In this mode, a fraction of the fibres are allocated to sky
positions, and their outputs median combined to form a single, high
S/N sky spectrum, which is then scaled and subtracted from each object
spectrum. The most efficient number of sky fibres to use in terms of
maximising the combined S/N of all object spectra, for sky limited
observations, is N$_{\rm{sky}}$ = 1.25 $\sqrt{\rm{N}_{\rm{fibres}}}$

\begin{figure}
  \begin{center}
    \begin{tabular}{c}
      \includegraphics[height=9cm]{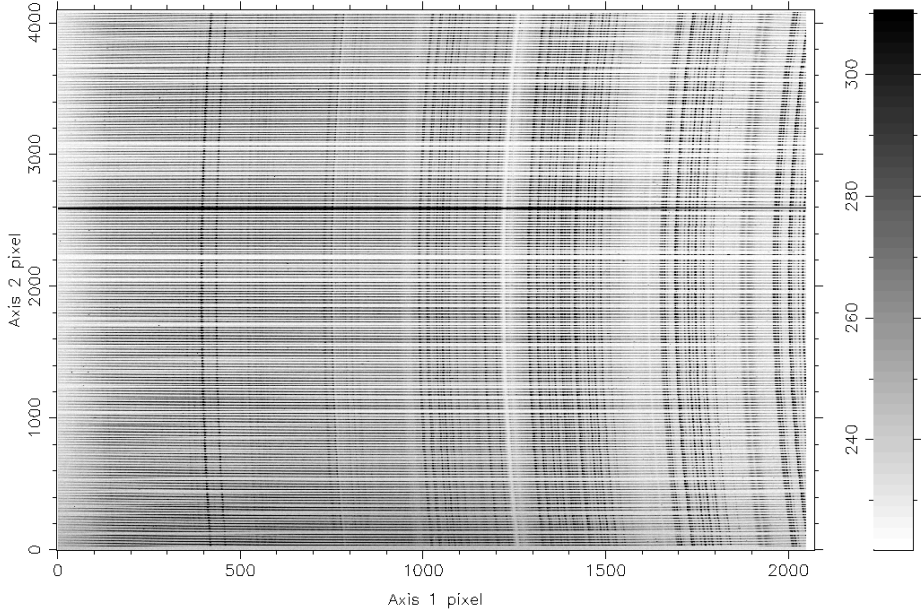}
    \end{tabular}
  \end{center}
  \caption{\label{fig:2D} This 2hour (4$\times$1800sec) raw AAOmega
  frame shows the full 2D CCD frame containing the 392 science fibre
  spectra.  Dispersion runs left to right in the low resolution red
  ($\lambda_{\rm{cen}}$=7250\AA, R$\sim$1300) unextracted spectra.
  Note the spectral curvature in the unsubtracted sky emission lines.}
\end{figure}

\begin{figure}
  \begin{center}
    \begin{tabular}{c}
      \includegraphics[height=9cm]{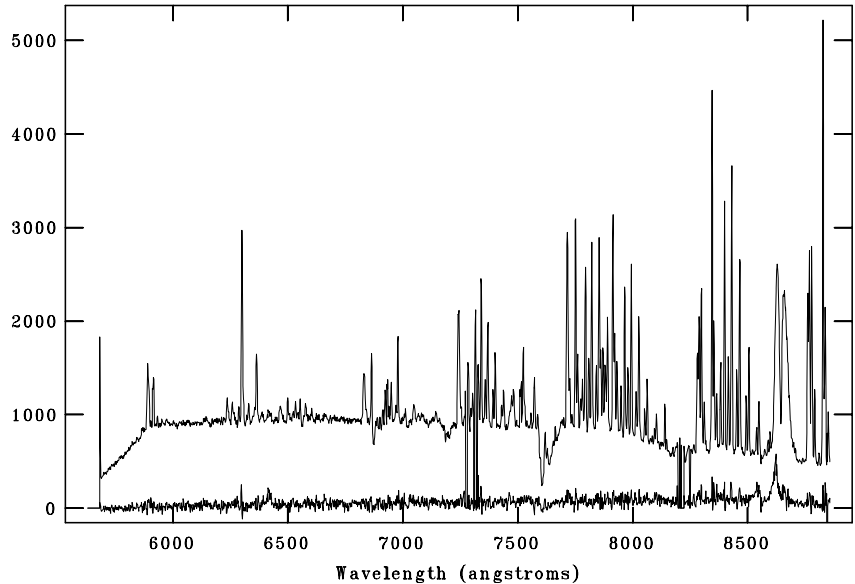}
    \end{tabular}
  \end{center}
  \caption{\label{fig:Skysub} An example spectrum shows the $\sim$1\%
  sky subtraction which can be routinely achieved with AAOmega.  The
  lower trace shows the spectrum of the underlying faint quasar target
  once the strong night sky spectrum (upper trace) has been
  subtracted.  Bad pixel masking has not yet been full integrated into
  the reduction software, as evidenced by the two regions of poor data
  quality.  The well sampled PSF allows most lost information to be
  recovered via simple interpolation.}
\end{figure}

The matching of sky to object spectra can be done either from the
median strength of strong sky emission lines, or from separate offset
sky or twilight sky frames.  Overall, we are aiming for 1\% precision
for the sky subtraction in this mode, which matches the Poisson noise
in the strongest sky lines for exposures of about an hour. This is
achieved routinely but not yet consistently, due to residual PSF
variations and spectrophotometric errors.  Sky subtraction accuracy,
using dedicated sky fibres,is demonstrated in figures \ref{fig:2D} and
\ref{fig:Skysub}.

For observations requiring more accurate sky subtraction, we use
Nod\&Shuffle (N\&S) sky subtraction$^{\rm{[5]}}$. In this mode, the
telescope is nodded between targets and sky, on a $\sim1$ minute
timescale; simultaneously, the spectra are charge-shuffled up and down
the detector, so as to build up separate sky and object spectra. In
this way, we observe the sky through exactly the same fibres and over
the same timescale as the objects.

The simplest variant of this mode is called \emph{mini-shuffling}; in
this mode, spectra are shuffled a few pixels, to produce minimally
resolved sky and object spectra. This allows all fibres to be used,
and tests suggest a sky subtraction accuracy of 0.3\% can be
achieved. However, there is a throughput hit of a factor 2 compared
with dedicated sky fibres, because of the extra sky noise and reduced
time on target. Mini-shuffling has not yet been fully commissioned.

\emph{Classic} N\&S involves using only half the fibres, and shuffling
into the vacated spaces on the detector. Sky subtraction is good to at
least the 0.1\% level; the achievable level has never yet been tested,
since even multi-night observations are now Poisson-limited in their
sky-subtraction.

Nod\&Shuffle can be combined with Cross-Beam Switching (CBS), where
two fibres are allocated to each source, with fixed physical
separation; the telescope is nodded back and forth to put the light
down each set of fibres in turn. This again halves the number of
objects that can be observed, but ensures that each is observed
continuously.

\section{Example proposed large science programs}
\aaomega, with both its MOS and SPIRAL feeds, is in high demand as a
general user facility at the AAT, accounting for $\sim$67\% of the
requested time in Semester 2006B (the second in which it was offered,
and the first proposal round after first light).  However, perhaps the
best way to showcase the capabilities of \aaomega\ is to examine a
subsection of the application for large programs which it has
engendered.  What follows is a brief, and incomplete, summary of some
such projects with which the instrument team has some familiarity.  At
the time of writing (April-May 2006) the decision of the time
assignment committee as to which programs to undertake, was still
pending.

\subsection{Dark Energy experiments}
Since being declared by much of the popular scientific press as one of
the most important scientific questions of the new millennium many
different strategies have been proposed to determine the nature of the
Dark Energy.  The success of the Baryon Oscillations measurements,
independently determined by the 2dF Galaxy Redshift Survey
(2dFGRS$^{\rm{[7,8]}}$) and the SDSS Luminous Red Galaxy program
(LRG$^{\rm{[9]}}$) has naturally given voice to the possibility of
using \aaomega\ to make the next step in such measurements.  Measuring
baryon oscillations at a higher redshift than current detections
provides a lever arm with which to constrain the evolution of dark
energy.  Two such programs currently under consideration by AATAC,
both of which promise their first results constraining dark energy on
the 2-3year time frame, are:\\
\newline
``WiggleZ'' - Drinkwater \etal\\
\newline
``The \aaomega\ Luminous Red Galaxy Redshift Survey'' - Croom \etal.\\
\newline
While both surveys target galaxies over the redshift range
0.5$<$z$<$0.9, \aaomega-LRG will target luminous red galaxies in the
same manner as the previous SDSS-LRG and 2SLAQ surveys, relying on the
many continuum features and breaks in LRG spectra to determine a
redshift, while WiggleZ will use a low redshift implementation of the
\emph{Lyman break} selection technique$^{\rm{[10]}}$, using imaging
data in the UV from the orbiting \textsc{galex} satellite mission, to
select emission line galaxies.  WiggleZ is a high target density,
\emph{small} area study, while \aaomega-LRG would cover
4,500\,deg$^{\rm2}$, using the \textsc{eso vst-atlas} survey (Shanks
\etal) as its primary input catalogue.

For success, both programs require the high sensitivity available with
\aaomega, particularly in the redder wavelengths, and 1\% sky
subtraction.  Both will require around 150-250 nights of telescope
time over the next 3-4 years to achieve their goals.

\subsection{Stellar surveys}
While many highly successful stellar population and radial velocity
studies were undertaken with the 2dF system, the improved stability
and higher resolution offered by \aaomega\ have generated a number of
exciting proposals for galactic or galaxy structure astronomy.

The AAOmagellan project (van Loon \etal) aims to \emph{``provide
radial velocities and abundance estimates for 300,000 stars, sampling
all structural, kinematical, chronological and chemical stellar
populations in both Magellanic Clouds and the Magellanic Bridge.''}

The ARGUS project (Lewis \etal) proposes \emph{``a comprehensive
investigation of the fossil evidence of the processes shaping our
Milky Way Galaxy that is unprecedented in its scope.  The Galactic
survey will determine the kinematics and chemical compositions of some
250,000 stars located in the bulge/central bar, inner disk/spiral arms
and outer thin/thick disks.  This near-field cosmology approach will
enable a large number of outstanding fundamental issues in galaxy
evolution to be addressed''}

\section{Performance and use}
\aaomega\ was first offered to the community for semester 2006A and
began routine operations in February 2006.  While for the most part
commissioning was smooth and uneventful, culminating in eight nights
of high quality Science Verification (SV) data, a number of
outstanding issues and \emph{lessons learned} remain to be addressed.

\subsection{Overall instrument performance}
Commissioning observations indicate that the red arm is currently
performing at 75\% of the expected sensitivity (21\% peak efficiency).
At time of writing, the blue arm however appears to only achieve 50\%
of expectations (16\% peak efficiency). It is believed that a
$\sim$20\% sensitivity gain will be achieved by correcting the known
outstanding issues outlined below.

\subsection{Fibre positioning and seeing quality}
The sensitivity estimates for \aaomega\ are based on the premise that
light from a source can be accurately coupled to each science fibre.
The effects of systematic and random error in the field acquisition
process are considered in detail elsewhere$^{\rm{[11]}}$.  Any measure
of the performance of \aaomega\ must factor in the effects of fibre
misplacement, either directly attributable to the positioner accuracy
(current positioner tolerance is $<$20$\mu$m, $<$0.3\,arcsec), or
indirectly via errors in the target astrometry or the
sky-to-field-plate astrometric transform.  During \aaomega\
commissioning it has been confirmed that the current 2dF
implementation of this astrometric transform is inaccurate due to
mechanical deformation of the gripper gantry at the 50-100$\mu$m
level.  Indirect mapping of transformation of each field plate to 2dF
positioner internal X/Y positions is to be undertaken.  The mapping
will utilise a high accuracy illuminated graticule to derive the
transformation.  The graticule will be milled with a regular grid of
back illuminated fiducial holes, at a 20mm pitch, to allow an accurate
survey of the gripper gantry deformation.

\subsection{Grating characterisation}
The VPH grating set, detailed in table \ref{tab:grat}, delivered by
Ralcon Development Lab, has proved to be of a high quality.  However,
the fabrication process means that each grating does depart somewhat
from its initial specification (in both blaze angle and number of
lines per millimetre).  A full calibration of each grating is required
as each is brought into use.  This process is currently 60\% complete.

\begin{table}[h]
\caption{AAOmega grating choices.  To aid the astronomer not familiar
with the operation of VPH gratings, a grating calculators has been
made available on-line:
http://www.aao.gov.au/cgi-bin/aaomega\_calc.cgi}
\label{tab:grat}
\begin{center}       
\begin{tabular}{|l|l||l|l||l|l|l|}
\hline
\textbf{Grating} & \textbf{Blaze} & \textbf{Useful range$^{\dagger}$}
& \textbf{Coverage$^{\dagger}$} & \textbf{Angle}  & \textbf{Dispersion}
& \textbf{MOS Resolution$^{\ddagger}$}\\
              & \textbf{(nm)}    & \textbf{(nm)}               
& \textbf{(nm)}       & \textbf{(Degrees)}& \textbf{(nm/pix)}   
& \\
\hline
\hline
580V  & 450 & 370--580 & 210 & 8        & 0.1   & 1300\\
385R  & 700 & 560--880 & 320 & 8        & 0.16  & 1300\\
\hline
1700B & 400 & 370--450 & 65  & 18       & 0.033 & 3500\\
1500V & 475 & 425--600 & 75  & 20--25   & 0.037 & 3700\\
1000R & 675 & 550--800 & 110 & 18--22.5 & 0.057 & 3400\\
1000I & 875 & 800--950 & 110 & 22.5--25 & 0.057 & 4400\\
\hline
3200B & 400 & 360--450 & 25  & 37.5--45 & 0.014 & 8000\\
2500V & 500 & 450--580 & 35  & 37.5--45 & 0.018 & 8000\\
2000R & 650 & 580--725 & 45  & 37.5--45 & 0.023 & 8000\\
1700I & 860 & 725--900 & 50  & 37.5--45 & 0.028 & 8000\\
\hline
1700D & 860 & 845--870 & 40  & 47$^{*}$ & 0.024 & 10000\\
\hline
\end{tabular}
\newline
\end{center}
$^{\dagger}$\,The useful wavelength range for the gratings, in
particular for the high-res gratings, is larger than the detector
width (2K in the dispersion axis).  Hence the spectrograph set-up must
be tuned to the required wavelength and Blaze.
\newline
$^{\ddagger}$\,IFU resolution is somewhat higher, due to the smaller
projected fibre diameter
\newline
$^*$\, The 1700D Ca\,\textsc{iii} grating is a tuned to give maximum
performance at 860nm but is unsuitable for observations at other angles.
\end{table}

\subsection{Pupil misalignment}
On integration of the spectrograph and the 2dF positioner, Hartmann
shutter focus tests readily demonstrated a minor spectrograph pupil
misalignment.  Investigations have traced the problem to an unforeseen
foible of the manufacturing process for the sub slits that comprise
the individual elements of the full \aaomega\ slit.  The support
mechanism used to hold the fibres in place within each slit-let, as
they are glued into place, has resulted in a depression of the fibres
at the end of each slit-let and hence has introduced an error in the
launch angle of light from the polished fibres ends.

The medium term solution is a re-shimming of each slit-let installed
in the slit to correct the launch angle.  Longer term solutions
include adjusting the slit unit field lens to steer the beam, or
\emph{simply} re-terminating the 90 slit-lets that comprise the full
\aaomega\ system.  Slanting the slit vain is undesirable due to the
increased vignetting this would introduce.

\subsection{Blackening of the slit unit faces}
An unfortunate side effect of the polishing process required to
prepare the slit units is that each slit-let is left with a highly
reflective polished metal surface in the beam and only $\sim$10s of
microns below the science fibre exit faces.  These surfaces can give
rise to back reflections and scattered light.  The effect is most
noticeable in arc frames, due to their high contrast data, and is a
low level effect.  However, to retain the high fidelity of \aaomega\
data the front of each slit unit will be blackened to suppress
reflections.

\subsection{Light inside \aaomega}
An unfortunate side effect of the ready availability of small, cheap,
low power light sources, in a bewildering array of colours, means it
now proves virtually impossible to source electronic equipment without
status indicator lights.  Ensuring that such illumination is bypassed,
deactivated or otherwise extinguished has required considerable
effort.  Limit and proximity switches are noted as the worst
offenders.  We also require that all optical encoders are deactivated,
all encoded motorised actuators de-energised and all pneumatic valve
state indicators are suitably light sealed during observations.

That this is essential becomes obvious when one realises that these
systems are no longer just inside the dome but actually inside
one's spectrograph!  \aaomega\ has been designed from the start
with this in mind, but a consequence of the excellent scattered light
suppression, and high sensitivity, is that every last light leak must
ultimately be tracked down and eliminated.

\subsection{Fringing}
At intervals during its working life a varying level of \emph{fringing
fibres} was seen in 2dF observations.  These interference fringes were
attributed to fractures within fibres, which resulted in air-gaps akin
to micro etalons.  Fringing occurred in two flavours, a stable mode
which could be removed via a flat field, and an unstable mode which
could not.  Typically unstable fringing indicated a fibre which would
soon break or a fibre button which was about to shed its prism.
Despite efforts to avoid the recurrence of fringing in \aaomega\
fibres, a substantial number of \aaomega\ fibres currently exhibit
unstable fringing.  Detailed analysis shows that the fringing
stabilises with time but instability is provoked by positioning the
damaged fibres.  A detailed investigation of the characteristics of
the fringing suggests that the fracture resides between the front face
of the prism at prime focus and the anchor point for the fibre at the
bottom of its 2dF retractor and not within the body of the 38m fibre
cable, or within \aaomega.

At this time it is believed that the root cause of the fibre
fracturing, which appeared mostly between December 2005 and January
2006 during an extreme heat wave, and which has been observed to be
stable since mid January 2006, is differential expansion of the steel
ferrule and the fibre at the fibre button, which has introduced an air
gap between prism and fibre while still leaving the prism securely
attached to its ferrule.  Further investigation is under way but if
confirmed, the correct solution is the relatively simple task of
re-terminating the affected fibres, an operation which would be
performed as a matter of course during the life of \aaomega\ to repair
the occasional damaged fibre.

\subsection{Blue arm sensitivity}
The reason for the reduced sensitivity of the blue arm (currently
running at $\sim$50\%) is unclear.  Correction of the minor problems
listed in this section will account for a 10-20\% gain.  Further
investigations are under way at this time to determine the source of
the efficiency loss.  Likely culprits are an incorrect coating on an
optical surface, poor CCD sensitivity or poor dichroic performance on
reflection.

\subsection{Arc lamp calibration system}
The current calibration system is based on the original 2dF system,
whereby the observing field plate is illuminated by arc and flat lamps
mounted in the top end chimney via two deployable flaps which are
placed into the top end beam.  This system allows wavelength
calibration from a combination of FeAr, CuAr, ThAr and CuNe hollow
cathode lamps.  Results are currently measured to better than a tenth
of a pixel.  Detailed investigations show that this limit to the
calibration accuracy, which impacts strongly on the accuracy of sky
subtraction when using dedicated sky fibres (see section
\ref{sec:sky}), is governed primarily by the difference in
$\emph{f}$-ratio between the arc lamp illumination and science
observations.  This issue was suspected with the 2dF system but was
not a measurable defect until the advent of \aaomega.  A program to
implement an upgraded calibration system, whereby the telescope pupil
is point sampled with arc light, probably by a fibre fed system, is
under way.  Observations at redder wavelengths, where sufficient night
sky OH air-glow lines are found (such as the Calcium Triplet region
typically used to study stellar radial velocities) are not degraded by
the use of the current system since the OH lines are used as a
secondary calibrator during data reduction.

\subsection{Point Spread Function variations}
Variations in the full field Point Spread Function (PSF) contribute
enormously to poor sky subtraction and were a significant defect in
the 2dF spectrograph system.  From the start \aaomega\ has been
designed around achieving a stable PSF as a function of wavelength.
The observed instrument PSF is found to be surprisingly Gaussian in
character.  Full Width Half Maximum (FWHM) varies slightly with
wavelength, as expected, and from grating to grating (due to
variations in grating character) but is satisfactory and stable at
$\sim$3.4 pixels.  This quality PSF is a key factor in achieving the
1\% sky subtraction often seen in \aaomega\ data to date.
Instabilities occasionally seen in the quality of sky subtraction may
be due to subtle departures from a uniform PSF across the entire CCD
field, and may require a PSF matching routine to achieve AAOmega's
ultimate sky subtraction performance, a task that is made
significantly easier from a high quality starting point.

\subsection{\label{sec:frost} Camera frosting}
Frosting has been observed on the coldest surfaces (including the
field flatteners). This is due to ``moisture'' adsorbed on the large
surface area of internal surfaces during camera fabrication.  When the
cameras are evacuated and cooled this slowly migrates to the coldest
surfaces and forms the observed frost.  This frosting has been
dropping with time, with the aid of a regime of dry nitrogen purging
and thermally cycling, and is now minimal.  The level will be
monitored over the coming months to assess the ultimate level that
will be achievable.

\section{Field preparation and data reduction}
An essential part of the instrument is comprehensive software for
preparing fields and reducing data.  The 2dF field configuration
software package has been updated to cater for AAOmega, and a new
version is being tested which uses simulated
annealing$^{\rm{[12,13]}}$ to find the most efficient allocation of
fibres to targets.

The 2dFDR data reduction package has been comprehensively overhauled,
with many improvements and new features added:

\begin{itemize}
\item A secondary wavelength calibration off sky lines in the
extracted spectra is now implemented; this greatly improves both sky
subtraction and radial velocity precision
\item Tramline fitting and spectrum identification is now fully
automated using pattern matching based on the eight missing spectra
where the guide fibres would be 
\item Splicing of low dispersion red and blue spectra is now
incorporated 
\item The software now works on \texttt{fits} files without conversion to
\texttt{ndf} format
\item Data for partially overlapping fields can now be seamlessly
combined on an object-by-object basis 
\end{itemize}

Further improvements are scheduled, including crude spectrophotometric
calibration from flux standards and improved self-consistent scattered
light subtraction.

\section{Future developments}
\aaomega\ began routine operations in February 2006.  A small number
of outstanding sub-systems still require integration and a short
program of simple upgrades, suggested during commissioning, will be
undertaken over the next six months with the goal of recovering the
full \aaomega\ design specification.  In the longer term the AAO is
currently considering the upgrade path for \aaomega.  The success of
the 2dF system and the broad utility of the \aaomega\ design suggests
the possibility of an infrared upgrade to \aaomega.  The proposed
project, AAOmicron, is currently in the early stages of specification,
with the intention to move to a full feasibility study in the later
half of 2006.  AAOmicron would reuse the 2dF wide field corrector and
ADC (providing close to the full 2\,degree field of view) and
\aaomega\ fibre runs, all of which have been fortuitously constructed
from materials which allow operation in the \emph{ZJ} and \emph{H} IR
bands.  AAOmicron would operate at wavelengths short of 1.8$\mu$m due
to the emissivity of the \aaomega\ fibres at longer wavelengths.  A
simplified, unarticulated copy of the \aaomega\ collimator and camera
system would be placed inside a -50$^{\circ}$ cold room to suppress
thermal background.  The use of a Rockwell Hawaii2 IR array,
preferably using the new \emph{K} blind technology making the chip
insensitive to thermal emission long ward of 1.7-1.8$\mu$m would allow
the camera to operate at higher temperature than typical for a
\emph{K} band system, and allow between 100-200 \aaomega\ fibres to be
projected onto the detector at one time using an only slightly modifed
AAOmega camera design.  With a field of view $\times$16 that of the
Subaru-FMOS system, and a fibre aperture more closely matched to the
typical scale length of redshift z$<$1.0 galaxies, the AAOmicron
system would represent an interesting complement to currently
available IR spectroscopy facilities, and would be the survey
instrument of choice in a number of fields.

The AAOmicron will be presented to the Australian community for
consideration in May 2006 and, if supported, will proceed to the
design study phase in the second half of 2006.  It is envisaged that
AAOmicron could be in operation by late 2010.





\section*{References} 

\noindent
[1]I.J.Lewis \etal, ``The Anglo-Australian Observatory 2dF facility'',
MNRAS \textbf{333}, 279-299, 2002

\noindent
[2]I.J.Lewis, K.Glazebrook and K.Taylor, ``Anglo-Australian Observatory
2dF project: a status report after the first year of scientific
operation'', \emph{Optical Astronomical Instrumentation}, ed. Sandro
D'Odorico, Proc. SPIE \textbf{3355}, 828-833, 1998

\noindent
[3] G.Smith \etal, ``AAOmega: a multipurpose fiber-fed spectrograph for
the AAT'' \emph{Ground-based Instrumentation for Astronomy},
ed. A.F.M.Moorwood \& I.Masanori, Proc. SPIE \textbf{5492}, 410-420,
2004

\noindent
[4] W.Saunders \etal, ``AAOmega: a scientific and optical overview'',
\emph{Ground-based Instrumentation for Astronomy}, ed. A.F.M.Moorwood
\& I.Masanori, Proc. SPIE \textbf{5492} 389-400, 2004

\noindent
[5] K.Glazebrook and J.Bland-Hawthorn, ``Microslit Nod-Shuffle
Spectroscopy: A Technique for Achieving Very High Densities of
Spectra'', PASP \textbf{113 issue 780}, 197-214, 2001

\noindent
[6] S.C.Barden, J.A.Arns and W.S.Colburn, ``Volume-phase holographic
gratings and their potential for astronomical applications'',
\emph{Optical Astronomical Instrumentation}, ed. S.D'Odorico
Proc. SPIE \textbf{3355}, 866-876, 1998

\noindent
[7] W.J.Percival \etal, ``The 2dF Galaxy Redshift Survey: the power
spectrum and the matter content of the Universe'', MNRAS \textbf{327
issue 4}, 1297-1306, 2001
	
\noindent
[8] S.Cole \etal, ``The 2dF Galaxy Redshift Survey: power-spectrum
analysis of the final data set and cosmological implications'', MNRAS
\textbf{362 issue 2}, 505-534, 2005

\noindent
[9] D.J.Eisenstein \etal, ``Detection of the Baryon Acoustic Peak in
the Large-Scale Correlation Function of SDSS Luminous Red Galaxies'',
ApJ \textbf{633}, 560-574, 2005

\noindent
[10] Steidel and Hamilton, ``Deep imaging of high redshift QSO fields
below the Lyman limit. I - The field of Q0000-263 and galaxies at Z =
3.4'', AJ \textbf{104 no.3}, 941-949, 1992

\noindent
[11] P.R.Newman, ``Positioning Errors and Efficiency in Fiber
Spectrographs'', PASP \textbf{114 issue 798}, 918-928, 2002.

\noindent
[12] W.H.Press \etal, \emph{Numerical Recipies in Fortran 77: the art
of scientific computing}, 2nd ed., Chap 10.9, Cambridge University
Press, Cambridge, 1992

\noindent
[13] B.Miszalsi \etal,``Multi-Object Spectroscopy Field Configuration
by Simulated Annealing'', MNRAS, 2006 submitted

\end{document}